\documentstyle[prb,aps,epsfig]{revtex}
\begin{document}
\draft
\preprint{HEP/123-qed}
\title{Josephson effects in a superconductor-normal metal mesoscopic structure with a dangling
superconducting arm.\\}

\author{R.Shaikhaidarov$^{+}$, A.F.Volkov$^{\ast \dagger }$ , H.Takayanagi$^{+}$, V.T.Petrashov$
^{\star }$, P.Delsing$^{\ddagger }$}
\address{$^{+}$NTT Basic Research Laboratories, \\
3-1 Wakamiya, Morinosato, Atsugi-shi,Kanagawa 243-0198, Japan \\
$^*$Theoretische Physik III,
Ruhr-Universitaet Bochum\\
Universitaetsstrasse 150,
D-44780 Bochum, Germany\\
$^{\dagger}$Institute of Radioengineering and Electronics of the Russian \\
Academy of Sciencies, Mokhovaya str.11, Moscow 103907, Russia.\\
$^{\star }$Royal Holloway University of London\\
Egham,Surrey TW20 0EX,U.K.\\
$^{\ddagger }$Department of Microelectronics and Nanoscience, Chalmers University of Technology\\
SE-412 96 Goteborg, Sweden}
\date{\today}
\maketitle

\begin{abstract}
We studied a mesoscopic cross-like normal metal structure
connected to two superconducting (S) and two normal (N)
reservoirs. We observed the Josephson effect under unusual
conditions when there is no current through one of the two S/N
interfaces. The potential difference between the S reservoirs was
zero unless the voltage applied between S and N reservoirs
exceeded a critical value although the electric potential in the
N wire connecting the superconductors varied
 in a nonmonotonic
way. The observed effects are discussed theoretically.
\end{abstract}

\pacs{PACS number: 74.50.+r, 73.23.-b, 85.25.-j}

\narrowtext
\twocolumn


The first studies of transport in superconductor-normal metal
(S/N) mesoscopic structures were focused on the dependence of the
normal metal conductance $G_{N}$ on the phase difference
$\varphi$ between the superconductors S. It was established that
$G_{N}$ is a periodic function of the phase difference $\varphi$
\cite{r1,r2,r3,r4,r5,r6}. The other problem - the effect of a
current in the normal wire on the critical Josephson current
$I_{c}$ (see Fig.1) - has been studied recently. Amongst other
findings obtained in the course of these studies there are two
remarkable effects. The first one is the so-called sign reversal
effect. The critical current $I_{c}$ changes sign in the
structure similar to the one shown in Fig.1 if the current
flowing between the N reservoirs exceeds a certain value. This
effect was studied in a Nb/Au mesoscopic structure with a short
mean free path (diffusive regime) \cite{r7}. The sign reversal
effect was analyzed theoretically in Refs. \cite{r8,r9,r10}
(ballistic regime) and in Refs. \cite{r11,r12,r13,r14} (diffusive
regime). Another interesting effect was observed in a diffusive
Al/GaAs mesoscopic structure \cite{r15}. It was found that an
additional current driven through the doped semiconductor GaAs
results in a non-monotonic behavior of the critical current in
the Al/GaAs/Al Josephson junction $I_{c}$. The current $I_{c}$
first decreases with increasing $V_{N}$, then increases and
reaches a maximum value $I_{m}$ when the voltage between the
semiconductor and superconductors  $V_{N}$ is of the order
$\Delta/e$. This effect was analyzed theoretically in Refs.
\cite{r12,r16}.

In both cases mentioned above an additional current was passed
through the normal wire. However the critical current was measured
as the critical current in an ordinary Josephson S/N/S junction;
i.e. as a maximum current flowing through both S/N interfaces at
zero voltage between the superconductors. It turns out that
Josephson-like effects can be observed in multiterminal
structures under rather unusual conditions when there is no
current through one S/N interface. Consider the structure shown
in the Fig. 1. Reservoirs S' and one N' are disconnected from the
external circuit and the current flows from the right N reservoir
to the upper S reservoir. In this structure Josephson-like
effects also arise. The prediction was made in Ref. \cite{r11} and
briefly discussed in Ref. \cite{r17}, but up to now it was not
observed experimentally. In this paper we report on experimental
studies of the effect, discuss its physical nature and present
results of theoretical analysis.

 First we discuss the physics of the effects using a simple phenomenological model. Later the
 main features of this model will be reproduced on the basis of a microscopic approach. For
 simplicity we consider a structure similar to that shown in Fig.1 in which the left N'
 reservoir is absent. The currents in the normal wires can be written as follows
\begin{equation}
{I_{1,2}}={I_S}+{I_{qp1,2}}
\end{equation}
\begin{equation}
I=(V_{0}-V_{N})/R_{h}(\varphi )
\end{equation}
Here $I_{S}=I_{c}\sin\varphi$ is the supercurrent in which the
critical current is a function of the electric potential $V_{N}$,
the second term in Eq.(1) is the quasiparticle current:
$I_{qp1,2}=\pm(V_{S,S'}-V_{0})/R_{1,2}(\varphi)$, $V_{0}$ is the
potential at the crossing point, $V_{S,N}$ are the potentials at
the S and N reservoirs, the potential at the S' is set equal to
zero (the potential $V_{N}$ is negative if $V_{S}$ is positive).
The resistances of horizontal and vertical arms $R_{h}$,
$R_{1,2}$ are functions of the phase difference $\varphi$; in the
case of a weak proximity effect they can be represented in the
form $R_{1,2}(\varphi)=R_{1,2}-\delta{R_{1,2}\cos\varphi}$.
Consider first the dc effect when
$V_{S}=\hbar{\partial_{t}}\varphi/2e=0$. The current through the
dangling superconducting arm is zero; this means that
$V_{0}=-I_{c}{R_{2}\sin\varphi}$, i.e. the quasiparticle current
is compensated by the supercurrent. Excluding $V_{0}$ from Eqs.
(1-2) and assuming that $R_{1}=R_{2}$, we obtain
\begin{equation}
V_{N}=-{({R_{h}}(\varphi)+{R_{1}}(\varphi)/2)}\cdot2{I_{c}}(V_{N})\sin\varphi
\end{equation}
\begin{equation}
I=I_{1}=2I_{c}(V_{N})\sin\varphi
\end{equation}
Eq.(3) determines a relation between $\varphi$ and $V_{N}$ and
Eq.(4) describes the form of the I-V curve if the phase
difference $\varphi$ is expressed as a function of $V_{N}$.
Therefore the phase difference in this structure is not
arbitrary, but is governed by the voltage $V_{N}$. Particularly
in the case of the small voltage $V_{N}$, we obtain for the phase
difference $\varphi{\cong-V_{N}/(2I_{c}R_{0})}$ and for the
current  $I=-V_{N}/R_{0}$, where $R_{0}=(R_{h}+R_{1}/2)$. In the
limit of high voltages the resistance of the structure increases
to the value $(R_{h}+R_{1})$. The critical voltage $V_{cr}$ is
defined as a maximum voltage $\vert{V_{N}}\vert$ above which a
finite voltage arises between the superconductors. This value can
be found from Eq.(3) as a maximum $\vert{V_{N}}\vert$ for which a
solution of Eq.(3) exists. Then, as follows from Eq.(4), the
effective critical current $I_{cr}$ is equal to
\begin{equation}
I_{cr}=2I_{c}(V_{cr}){\sin\varphi}{(V_{cr})}
\end{equation}

If the voltage $V_{N}$ or the current $I$ exceeds the critical
value, one needs to solve Eqs.(1-2) taking into account a finite
voltage $V_{N}$ between the superconductors. These equations
cannot be reduced to a dynamic equation for the phase $\varphi$
which describes a single Josephson S/N/S junction. However in this
paper we will not discuss ac Josephson effects in the structure
under consideration.

The analysis of the situation, when a current $I$ is passed
between a normal reservoir and one of the superconductors, carried
out on the basis of this simple model shows the following
features. For $\vert{V_{N}}\vert<V_{cr}$ the potential difference
between superconductors remains zero, therefore a vertical line
on the $I(V_{S})$ curve should arise; a nonlinear part with a
finite slope on $I(V_{N})$ curve should appear. For
$\vert{V_{N}}\vert>V_{cr}$ a kink appears on the current-voltage
characteristics. This picture is confirmed by the present
experimental data and by a theoretical analysis carried out with
the help of a microscopic theory (quasiclassical Green's function
technique).

The sample geometry is shown schematically in Fig.1. The structure
we studied consists of two crossed, $50$ nm thick and $110$ nm
wide Ag wires with $50$ nm thick and $500$ nm wide Al leads
attached to the vertical wire and $350$ nm thick and $20$ ${\mu}$m
wide Ag reservoirs attached to the horizontal wire. Electron beam
lithography and lift-off technique were used to produce samples.
In order to ensure high transparency interfaces, we cleaned the Ag
films before the evaporation of the Al film via Ar sputtering.
The interface resistance was estimated to be of the order of the
normal state resistance of the sample. We determined the mean free
path $l=37$ nm and the diffusion constant $D=124$ cm$^2$/s from
the measured resistance of Ag wire. The phase breaking length
$L_{\varphi}=1.5$ ${\mu}$m was obtained from magnetoresistance
measurement of a coevaporated Ag wire at the base temperature.
The length of the normal part $L_{NN'}$ and the distance between
the superconductors $L_{SS'}$ were $1.3$ ${\mu}$m and $0.5$
${\mu}$m respectively. The coherence length of the normal metal is
equal to $\xi_{T}=\sqrt{\hbar\ D/k_{B}\ T}=1.3$ ${\mu}$m at the
base temperature $50$ mK.

We performed measurements as follows. The current $I$ was passed
between the normal reservoir N and superconducting reservoir S
(see Fig.1). The reservoirs N' and S' were not connected to the
measurement circuit. We measured the voltage $V_{1}$ between the
N' and S reservoirs and the voltage $V_{2}$ between the
superconductors. The results of measurements are presented in
Fig.2. The solid line represents Josephson-like effects in the
S/N/S structure with the dangling arm. The potential difference
between superconductors is equal to zero when the current is less
than critical $I_{cr}$ (solid line) despite the finite potential
difference between the crossing point and superconductors (dashed
line). As we mentioned before, the quasiparticle current $I$
splits into two currents $I_{qp1}$ and $I_{qp2}$ at the crossing
point flowing towards superconductors. The supercurrent $I_{s}$,
equal to the quasiparticle current $I_{qp2}$, flows between
superconductors in the opposite direction to $I_{qp2}$. In other
words, when the current $I$ is passed from N to S, the phase
difference adjusts in such a way that the potential difference
between the superconductors and net current through the dangling
arm are equal to zero. This means that the quasiparticle current
in the horizontal N wire creates the condensate current in the
vertical N wire. We would like to stress that contrary to the
case of the conventional dc Josephson effect, the electric
potential along the vertical N wire is not constant, but
decreases from a maximum at the crossing point to zero at the
superconductors due to the quasiparticle current. The dotted line
of Fig.2 represents the current-voltage characteristic of S/N/S'
junction measured in a conventional way, when current is passed
between superconductors (S and S') and voltage $V_{2}$ (see
Fig.1) is measured. The critical current $I_{c}\vert_{V_{N}=0}$
(dotted line) is less then $I_{cr}$ (solid line).

Here we present results of the analysis based on a microscopic
theory. This analysis qualitatively confirms the main features of
the phenomenological approach given above. We use the well
developed quasiclassical Green's function technique. (The
application of this technique to the study of transport in S/N
mesoscopic structures is reviewed, for example, in \cite{r18}.)
We consider the diffusive limit and restrict ourselves to the
consideration of the dc case ($V_{S}=0$). We solve equations for
distribution functions $f_{\pm}$ which describe the symmetric and
antisymmetric parts of population of the electron and hole-like
branches of the quasiparticle spectrum. In the one-dimensional
model assumed by us, these equations can be solved exactly and
formulae for the potential $V_{N}$ and current $I$ can be written
in terms of the retarded (advanced) Green's functions
$\hat{F}^{R(A)}$ which obey the Usadel equation. In the general
case these formulae are rather cumbersome. Here we present
expressions for the currents in the simplest case when the S/N
interface resistance $R_{S/N}$ is larger than the resistance of
the normal wire $R_{S/N}=1/(\sigma{L_{S}})$. In this case,
excluding $V_{0}$ we obtain for the current $I_{2}$.
\begin{equation}
I_{2}={I_{c}\sin\varphi+I_{+}+{I_{-}\cos\varphi}}
\end{equation}
Where $I_{2}$ is the current in the upper vertical arm. All the
functions depend on $V_{N}$ and have the form
\begin{equation}
I_{\pm}=(eR_{S/N})^{-1}{\int_{0}^{\infty}}{d\varepsilon}[{{Im({F_{S}}})}{{Im({F_{S}T_{\pm}/\theta_{S}})}}{f_{eq-}}],
\end{equation}
\begin{eqnarray}
I_{c}=&&(eR_{S/N})^{-1}{\int_{0}^{\infty}}{d\varepsilon}[{{Im({F_{S}}})}{{Re({F_{S}T_{-}/\theta_{S}})}}{f_{eq+}}\nonumber\\
&&
+{{Re({F_{S}}})}{{Im({F_{S}T_{-}/\theta_{S}})}}{f_{eq}}],
\end{eqnarray}
where
$F_{S}=\Delta/\sqrt{{(\varepsilon+i{\gamma})}^{2}-{\Delta}^{2}}$
is the retarded Green's function in the superconductors,
$T_{\pm}=\tanh\theta\pm\tanh\theta_{S}$,
$\theta=\theta_{S}+\theta_{N}$,
$\theta_{S,N}=\sqrt{-2i\varepsilon/\varepsilon_{S,N}}$,
$\varepsilon_{S,N}=D\hbar/L_{S,N}^{2}$ is the Thouless energy,
$f_{eq}=\tanh(\varepsilon/2k_{B}T)$,
$f_{eq\pm}=[\tanh((\varepsilon+eV_{N})/2k_{B}T)\pm\tanh((\varepsilon-eV_{N})/2k_{B}T)]/2$
are the distribution functions in the N reservoir. For the
current in the lower vertical arm we have the same expression
with opposite quasiparticle current
$I_{1}={I_{c}\sin\varphi-(I_{+}+{I_{-}\cos\varphi})}$. The
critical voltage is determined from the equation
$I_{1}=-I_{+}+\sqrt{I_{c}^{2}+I_{-}^{2}}\sin{(\varphi+\theta)}=0$
as a maximum value of $V_N$ for which a solution of this equation
for $\varphi$ exists (we note that $I_{+}$ increases with
increasing $V_{N}$), where
$\cos{\theta}=I_{c}/\sqrt{I_{c}^{2}+I_{-}^{2}}$. We find
$I_{+}(V_{cr})=\sqrt{I_{c}^{2}(V_{cr})+I_{-}^{2}(V_{cr})}$ and
the critical current is
\begin{equation}
I_{cr}=2I_{c}^{2}(V_{cr})/\sqrt{I_{c}^{2}(V_{cr})+I_{-}^{2}(V_{cr})}.
\end{equation}
The critical current corresponds to the phase difference $\varphi_{c}$ determined by the relation:
\begin{equation}
\sin\varphi_{c}=I_{c}(V_{cr})/\sqrt{I_{c}^{2}(V_{cr})+I_{-}^{2}(V_{cr})}.
\end{equation}
Fig.3 shows the experimental results and theoretical calculations
of temperature dependence for the critical current $I_{cr}$ and
the Josephson critical current $I_{c}\vert_{V_{N}=0}$. Where
$\gamma=0.1\Delta$, $\Delta(0)=1.76k_{B}T_{c}$, $T_{c}=1.4$ K,
$D=140$ cm$^2$/s and $R_{S/N}=3.75$ $\Omega$. The diffusion
coefficient and the interface resistance estimated from resistance
measurement were $124$ cm$^2$/s and $0.7$ $\Omega$ respectively.
The discrepancy of fitting parameters and estimated values can be
attributed to the fact that in our samples $R_{S}\cong{R_{S/N}}$,
while in our model interface resistance dominates over sample
resistance. Nevertheless the theoretical curves show
qualitatively the same behavior as the experimental ones. We note
that these formulae are a good approximation even if
$R_{S}\cong{R_{S/N}}$. We plot a guideline
$I^{\ast}=2I_{c}\vert_{V_{N}=0}$ for comparison. At temperatures
below $500$ mK, $I_{cr}$ deviates from $I^{\ast}$. There are two
reasons for that: first, the reduction of the critical current
$I_{cr}$ by the voltage $V_{N}$; second, the reduction of
resistances ${R_{h}}$, ${R_{1,2}}$ related to the proximity
effect. The latter means that the phase difference which
corresponds to $I_{cr}$ does not reach $\pi/2$
($\varphi_{c}<\pi/2$) (see eq. 3,10). At high temperatures the
curves coincide $I_{cr}\cong{I^{\ast}}$ since both corrections
are negligible.

 The S/N/S junction we have studied could be driven to the $\pi$-state by additional bias voltage
 applied between normal reservoirs N and N' (see  $V_{contr}$ of Fig.1). At a certain value of
 the bias, the current-phase
relationship changes in such a way that at zero current, phase
difference is equal to $\pi$ \cite{r7} due to the change in the
electron distribution function.
 It turns out that Josephson-like effects can be observed in an S/N/S junction driven to
 $\pi$-state,
 with no current through one S/N interface. First we performed
 measurements similar to ones reported in Ref. \cite{r7}. We measured the $I$-$V$ curves of S/N/S
 junction with additional bias voltage $V_{contr}$ applied between normal reservoirs N and N'.
 The dependence of critical current $I_{c}(V_{contr})$ was similar to the one observed by
 Baselmans et al. The critical current decreased with increasing $V_{contr}$ and disappeared
 at $V_{contr}^{\ast}=180$ ${\mu}$V. It reappeared again at higher voltages reaching a maximum value
 at $V_{contr}^{\ast\ast}=288$ ${\mu}$V (see Fig.4 circles). To ensure the change of phase-current
 relationship at $V_{contr}^{\ast}$ we measured the resistance of the horizontal wire depending on current
 flowing through the S/N/S junction. At additional bias voltages less than $V_{contr}^{\ast}$
 the resistance had a minimum at zero current, while at additional bias voltages more than
 $V_{contr}^{\ast}$ it had a maximum. We then performed measurements as follows. We passed a current
 between N and S measuring the potential difference between superconductors S and S'. Additional
 voltage bias $V_{contr}$ was applied between normal reservoirs N and N'. The dependence of
 the critical current $I_{cr}(V_{contr})$ was similar to $I_{c}(V_{contr})$ (See Fig.4 squares).
 We plot current-voltage characteristics at $V_{contr}^{\ast\ast}$ on the insert to Fig.4. We
 can conclude that despite different current-phase relationship for $\pi$-junction, the
 qualitative picture of current distribution remains the same: the quasiparticle current in
 the dangling arm is compensated by the supercurrent and the potential difference between superconductors
 is equal to zero.

In summary, we have studied the Josephson-like effects in an Al/Ag
mesoscopic structure with a dangling superconducting arm.
Experimental data on currents are in qualitative agreement with
theoretical results. The study of this effect may yield additional
information on the relaxation mechanism of the distribution
function and reveal new peculiarities on the $I(V_{N})$ curve at
voltages larger than the critical voltage $V_{cr}$.

We would like to thank V.Shumeiko for useful discussions. One of
the authors (AFV) is grateful to NEDO for the financial support
and to Prof. H.Takayanagi for hospitality. We acknowledge the
financial support of the NEDO International Joint Research Grant.
VTP acknowledges financial support from the EPSRC (Grant No.
GR/L94611).

\begin{figure}
\begin{center}
\epsfxsize=8cm \epsfbox{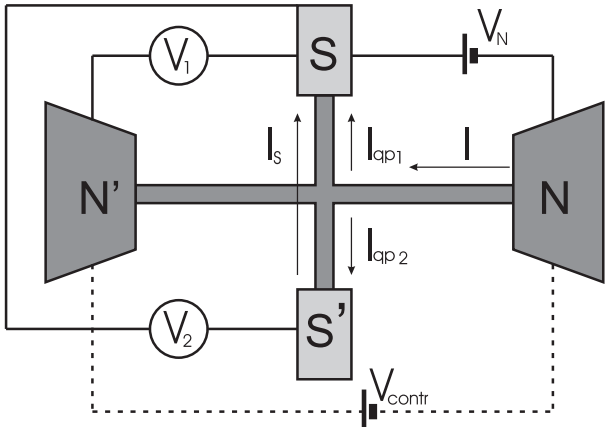} \caption{Sample and measurement
circuit schematic. S,S' and N,N' are superconducting and normal
reservoirs. Arrows show currents: $I=I_{qp1}+I_{s}$,
$I_{s}=-I_{qp2}$.}
\end{center}
\label{f1}
\end{figure}

\begin{figure}
\begin{center}
\epsfxsize=8.5cm \epsfbox{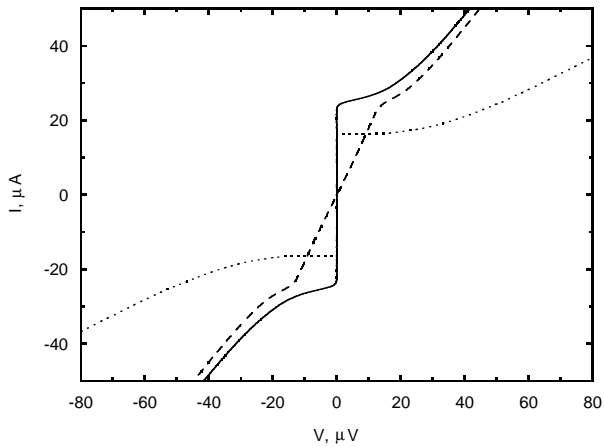} \caption{ Current-voltage
characteristics of S/N/S structure with dangling arm. Measurement
current is passed between reservoirs N and S for solid and dashed
lines. Solid line corresponds to the voltage $V_{2}$ measured
between superconductors S and S'. Dashed line corresponds to the
voltage $V_{1}$ measured between reservoirs N' and S (See Fig.1).
Dotted line represents an experimental $I$-$V$ curve of S/N/S'
junction.}
\end{center}
\label{f2}
\end{figure}

\begin{figure}
\begin{center}
\epsfxsize=8.5cm \epsfbox{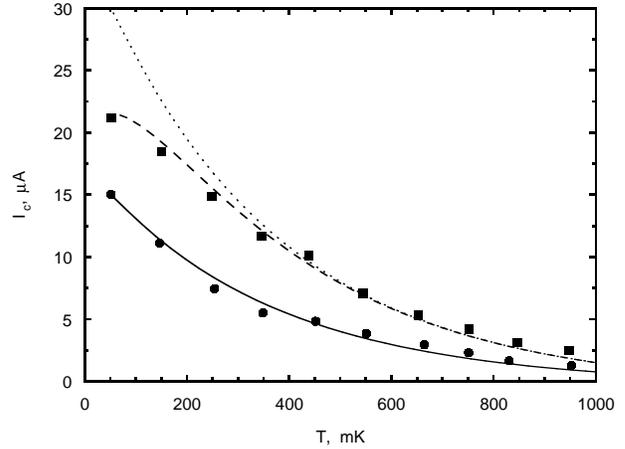} \caption{ Experimental and
theoretical temperature dependencies of critical currents.
Circles show the temperature dependence of the critical current
measured in a conventional way and squares show the same
dependence for the structure with a dangling arm. Calculated
temperature dependencies of critical currents are shown by lines.
Solid line is $I_{c}\vert_{V_{N}=0}$ ; dashed line is $I_{cr}$;
dotted line is guideline $I^{\ast}=2I_{c}\vert_{V_{N}=0}$.}
\end{center}
\label{f3}
\end{figure}

\begin{figure}
\begin{center}
\epsfxsize=8.5cm \epsfbox{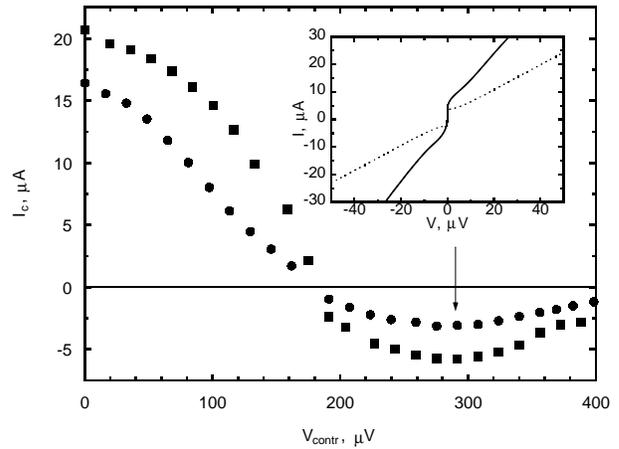} \caption{ Critical currents
$I_{c}$ and $I_{cr}$ versus additional bias voltage applied
between normal reservoirs measured at temperature $50$ mK. Circles
and squares show $I_{c}$ and $I_{cr}$ respectively. Inset:
current-voltage characteristics measured at additional bias
voltage shown by arrow. Dotted line represents the $I$-$V$ curve
of S/N/S' junction. Solid line corresponds to the measurement with
a dangling arm.}
\end{center}
\label{f4}
\end{figure}

\end{document}